\documentclass[5p]{elsarticle}

\usepackage{graphicx}
\usepackage{amssymb}
\usepackage{color}
\usepackage{colordvi}
\usepackage{ulem}

\biboptions{sort&compress}

\begin{document}
\begin{frontmatter}

\title{Constraining solar hidden photons using HPGe detector}


\author{R. Horvat}
\author{D. Kekez\corref{cor}}
\cortext[cor]{Corresponding author.}
\ead{Dalibor.Kekez@irb.hr}
\author{M. Kr\v{c}mar}
\author{Z. Kre\v{c}ak}
\author{A. Ljubi\v{c}i\'{c}}

\address{Rudjer Bo\v{s}kovi\'{c} Institute, P.O.Box 180, 10002 Zagreb,
             Croatia}
\date{\today}

\begin{abstract}
In this Letter we report on the results of our search for photons from a U(1) gauge
factor in the hidden sector of the full theory. With our experimental setup
we observe the single
spectrum in a HPGe detector arising as a result of the photoelectric-like absorption
of hidden photons emitted from the Sun on germanium atoms inside the
detector. The main ingredient of the theory used in our analysis, a severely
constrained kinetic mixing from the two U(1) gauge factors and massive
hidden photons, entails  both photon into hidden state oscillations and a
minuscule coupling of hidden photons to visible matter, of which the latter our experimental
setup has been designed to observe. On a theoretical side,
full account was taken of the effects of refraction and damping of photons
while propagating in Sun's interior
as well as in the detector.
We exclude hidden photons with kinetic couplings
$\chi > (2.2\times 10^{-13}-3\times 10^{-7})$ in the mass region 
$0.2~{\rm eV} \lesssim m_{\gamma^\prime} \lesssim 30~{\rm keV}$.
Our constraints on the mixing parameter $\chi$ in the mass region from 
$20~{\rm eV}$ up to $15~{\rm keV}$ 
prove even slightly better then those obtained recently 
by using data from the CAST experiment, albeit still  
somewhat weaker than those obtained from solar and HB stars
lifetime arguments.
\end{abstract}

\begin{keyword}
Hidden photon \sep Kinetic mixing \sep Sun 
\PACS 12.60.Cn \sep 14.70.Pw \sep 96.60.Vg 
\end{keyword}

\end{frontmatter}

The models of SUSY-breaking  most oftenly involve dynamics of a hidden
sector, which uses to communicate the SUSY-breaking scale (usually larger
than the weak scale) to the visible sector with aid of operators having always the
appropriate loop or Planck-scale suppression. Otherwise the gauge hierarchy
would be destabilized. If a U(1) gauge factor is contained in the hidden
sector, a new communication mechanism \cite{Dienes:1996zr} in the form of an operator that mixes
two U(1)'s opens up, having a
potential to destabilize any model for SUSY-breaking. Since this operator is
a renormalizable one, it comes with no suppression by the large mass scale
and therefore the mixing parameter $\chi$ must be small. The fact that both
in field theory and in string theory settings an appreciable amount of
$\chi$ can be generated \cite{Dienes:1996zr}, one may recognize the kinetic mixing operator as an
important ingredient in these fundamental theories.

Introduction of an explicit mass term for hidden photons (thereby not upsetting the
renormalizability of the theory) together with the
kinetic mixing term mentioned above would lead to a model of photon
oscillations (photons-hidden photons) \cite{Georgi:1983} similar to the much more popular
neutrino flavor oscillations. To this end, one gets rid of the kinetic
mixing term by the appropriate rotation of states, introducing in such a
manner a truly sterile state with respect to gauge interactions. This
generates a nondiagonal mass matrix in the sector of two photons, a
necessary ingredient for the oscillation phenomenon. A thorough analysis of
finding appropriate propagating states in vacua as well as in a matter
background has been done recently \cite{Redondo:2008aa,Redondo:2008ec}. Thus, the flavor (or
interacting) states (one
truly sterile while the other with the full gauge coupling to charged
matter particles) can be expressed as a linear combination of propagating states in
vacua/matter. As a consequence, a sterile propagating state would gain a
tiny coupling to ordinary matter of order $\chi$ in vacua, while in matter
such a coupling depends on both the real and imaginary part of the photon self-energy at
finite temperature/density. This is crucial for our experimental setup (see
below), since after being oscillated into a sterile state and (presumably) quick absorption of
the active component in ordinary matter, it is just the sterile propagating state that leaves
material background and travels unscathed towards a region where it is to be
detected. 

  In the present Letter, we aim to observe sterile photon states
(hereafter denoted as $\gamma^\prime$) in a few keV
range and coming from the Sun by observing the photoelectric-like process on
germanium atoms inside the HPGe detector. So far the most stringent limits
on the hidden-photon mixing, in the mass region relevant for our investigation,
are obtained by experiments using the Sun as a source of hidden photons
\cite{Redondo:2008aa} as well as by
astrophysical arguments regarding the solar lifetime \cite{Redondo:2008aa} 
and HB stars lifetime \cite{Jaeckel:2010ni}.
  
\begin{figure}[ht]
\centerline{\includegraphics[width=90mm,angle=0]{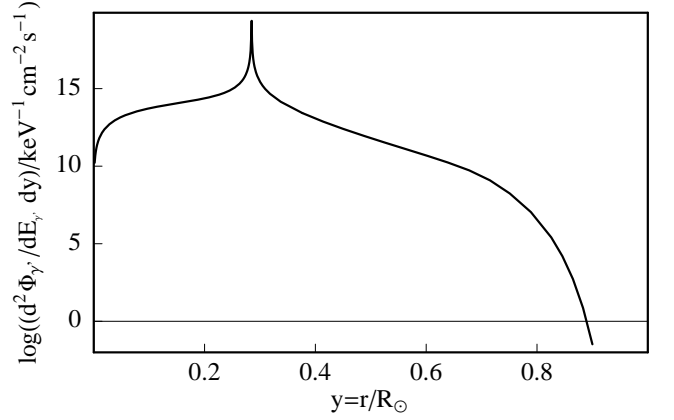}}
\caption{Flux of solar hidden photons at the Earth as a function of the
normalized radial coordinate $y=r/R_\odot$ for
$E_{\gamma^\prime}=1.6~\mbox{\rm\scriptsize keV}$,
$m_{\gamma^\prime}=100~\mbox{\rm\scriptsize eV}$, and $\chi=3\cdot10^{-13}$.
}
      \label{fluxe}
   \end{figure}

\begin{figure}[ht]
\centerline{\includegraphics[width=93mm,angle=0]{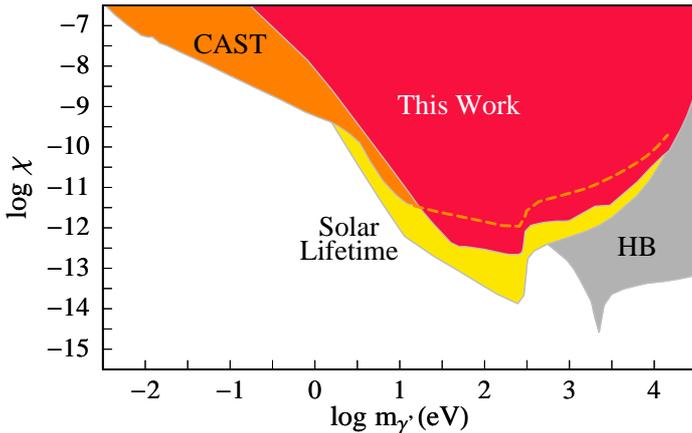}}
               \caption{Limits on the mixing parameter as a 
function of the hidden-photon mass from this experiment against
the current hidden photon bounds taken from 
\cite{Redondo:2008aa,Jaeckel:2010ni}. For description see the text.}
      \label{mixing}
   \end{figure}

The low-energy effective Lagrangian for the two-photon system with 
kinetic mixing reads \cite{Holdom:1985ag}
\begin{eqnarray}
{\cal L} 
&=&
-\frac{1}{4} F^{\mu\nu} F_{\mu\nu} 
-\frac{1}{4} F^{\prime\mu\nu} F^\prime_{\mu\nu}
-\frac{1}{2} \,\chi \,F^{\mu\nu} F^\prime_{\mu\nu}
\nonumber \\ &&
+\frac{1}{2} m_{\gamma^\prime}^2 A^{\prime\mu} A^\prime_\mu
-eA^\mu J_\mu~,
\label{HiddenPhotons:Lagrangian}
\end{eqnarray}
\noindent where $F^{\mu\nu}$ and $F^{\prime\mu\nu}$ are the photon ($A^\mu$) and
hidden photon ($A^{\prime\mu}$) field strengths, respectively, $J^\mu$ is the current
of electrically charged matter while $m_{\gamma^\prime}$ is the
hidden-photon mass that could arise from a Higgs or St\"{u}ckelberg
mechanism \cite{Stueckelberg:1900zz}. 
For transversely polarized hidden photons of energy sufficiently above 
the plasma frequency $\omega_{\rm p}$ (in the solar model, $1~{\rm eV} \lesssim \omega_{\rm p} 
\lesssim 295~{\rm eV}$) we can write the differential flux    
at the Earth as \cite{Redondo:2008aa}
\begin{eqnarray}
\frac{d\Phi_{\gamma^\prime}}{dE_{\gamma^\prime}}
&=&
\frac{1}{\pi^2 R_{\mbox{\rm\scriptsize Earth}}^2}
\int_0^{R_\odot}\, dr\, r^2\, 
\frac{E_{\gamma^\prime}\sqrt{E_{\gamma^\prime}^2-\omega_{\rm p }^2}}
     {e^{E_{\gamma^\prime}/(k_{\rm B}T)}-1}
    \nonumber \\
      && \times 
\frac{\chi^2 m_{\gamma^\prime}^4}
     {\left(\omega_{\rm p }^2-m_{\gamma^\prime}^2 \right)^2+
     (E_{\gamma^\prime}\Gamma)^2}\, \Gamma~,
\label{HiddenPhotons:flux:SI}
\end{eqnarray}
\noindent where $E_{\gamma^\prime}$ is the hidden-photon energy,
the plasma frequency  $\omega_{\rm p} = \sqrt{4\pi\alpha N_{\rm e}/m_{\rm e}}$,
$k_{\rm B}$ is the Boltzmann constant, $T$ is the solar plasma temperature, 
$R_\odot$ is the solar radius,
$R_{\mbox{\rm\scriptsize Earth}}\approx 1.5\times 10^{13}\,\mbox{\rm cm}$ is 
the average Sun--Earth distance, and $\Gamma$ is the damping factor given by 
 \cite{Redondo:2008aa}
\begin{eqnarray}
\Gamma
&=&
\frac{16\pi^2 \alpha^3}{3 m_{\rm e}^2 E_{\gamma^\prime}^3}
\sqrt{\frac{2\pi m_{\rm e}}{3 k_{\rm B} T}} N_{\rm e}
\left[1-\exp(-\frac{E_{\gamma^\prime}}{k_{\rm B} T})\right]
     \nonumber \\
      && \times 
\sum_{\rm i} Z_{\rm i}^2\, N_{\rm i}\, \bar{g}_{\mbox{\rm\scriptsize ff,i}} + 
\frac{8\pi\alpha^2}{3 m_{\rm e}^2}\, N_{\rm e}~.
\label{HiddenPhotons:DampingFactor}
\end{eqnarray}
\noindent The first term is the bremsstrahlung contribution to
the damping, where index ``i'' designates protons or alphas, while the second 
term is the Compton contribution. Here it is assumed that all hydrogen and helium are
completely ionized.  
The thermally averaged Gaunt factors $\bar{g}_{\mbox{\rm\scriptsize ff,i}}$ are taken from 
\cite{Itoh:1999wt} which presents an accurate analytic fitting formula for the 
nonrelativistic exact Gaunt factor. The calculation is also checked using another exact
formula (with numerical integration over Maxwellian distribution) \cite{Brussaard:1962zz}.
The $r$-dependent quantities, $T$, $N_{\rm e}$, $N_{\rm p}$, and $N_\alpha$ are calculated
using BS05 Standard Solar Model \cite{Bahcall:2004pz}. 
Our experiment is the most sensitive to hidden photons of around 1.6~keV
(see below), and since they are created mostly in Sun's inner layers (as shown in
Fig.~\ref{fluxe}), Eqs.~(\ref{HiddenPhotons:flux:SI}) and 
(\ref{HiddenPhotons:DampingFactor}) (ionization neglected)
can be reliably applied to calculate the expected flux of hidden photons.
The contribution of different solar layers to the hidden-photons flux,
depicted in Fig.~\ref{fluxe}  for $m_{\gamma^\prime}=100~{\rm eV}$,
exhibits a narrow peak at $r=0.28\, R_\odot$ corresponding to the
resonant contribution, $\omega_{\rm p}^2(r)=m_{\gamma^\prime}^2$, in the
integrand of Eq.~(\ref{HiddenPhotons:flux:SI}).
The plasmon mass is maximal at Sun's center,
$\omega_{\rm p}(0)\simeq 290$ eV. For $m_{\gamma^\prime}\gtrsim 290$ eV there
is no resonant contribution in Eq.~(\ref{HiddenPhotons:flux:SI})
(and there would have been no peak
in Fig.~\ref{fluxe} had it drawn for $m_{\gamma^\prime}\gtrsim 290$ eV).
This causes a sudden drop in sensitivity for
$m_{\gamma^\prime}\gtrsim 290$ eV what can be clearly seen
in Fig.~\ref{mixing}.

Our experiment involves searching for the particular energy spectrum in the measured data,
\begin{equation}
\frac{dN_{\gamma^\prime}}{dE_{\gamma^\prime}}
=
\frac{d\Phi_{\gamma^\prime}}{dE_{\gamma^\prime}}
\sigma_{\gamma^\prime {\mbox{\rm\scriptsize Ge}}\to 
{\mbox{\rm\scriptsize Ge}}^\ast {\rm e}}(E_{\gamma^\prime})
N_{\mbox{\rm\scriptsize Ge}}\, t~,
\label{signal}
\end{equation}
\noindent produced if the hidden photons from the Sun are detected via 
photoelectric-like effect on germanium atoms.
Here $N_{\mbox{\rm\scriptsize Ge}}$ is the number of germanium atoms in the detector and $t$
is the data collection time.
The cross section for the hidden-photon absorption,
$\gamma^\prime + {\mbox{\rm Ge}}\to {\mbox{\rm Ge}}^\ast + {\rm e}$,
can be expressed via the cross section for the ordinary photoelectric
absorption as (see, e.g., Ref.~\cite{Pospelov:2008jk})
\begin{equation}
\sigma_{\gamma^\prime {\mbox{\rm\scriptsize Ge}}\to {\mbox{\rm\scriptsize Ge}}^\ast {\rm e}}
       (E_{\gamma^\prime}) =
\frac{\chi_{\mbox{\rm\scriptsize eff}}^2}{\beta_{\gamma^\prime}}
\sigma_{\gamma {\mbox{\rm\scriptsize Ge}}\to {\mbox{\rm\scriptsize Ge}}^\ast {\rm e}}
       (E_{\gamma^\prime})~,
\label{crossSection}
\end{equation}
\noindent where
\begin{equation}
\chi^2_{\mbox{\rm\scriptsize eff}}
=
\chi^2
\left|
\frac{m_{\gamma^\prime}^2}
   {m_{\gamma^\prime}^2+2E_{\gamma^\prime}^2(n(E_{\gamma^\prime})-1)}
\right|^2~,
\end{equation}
and $\beta_{\gamma^\prime}=\sqrt{1-m_{\gamma^\prime}^2/E_{\gamma^\prime}^2}$
is the velocity of the hidden photons and the data for
$\sigma_{\gamma {\mbox{\rm\scriptsize Ge}}\to {\mbox{\rm\scriptsize Ge}}^\ast {\rm e}}$
are taken from Ref.~\cite{XCOM}.
The effective mixing parameter $\chi_{\mbox{\rm\scriptsize eff}}$ takes into
account the media (germanium) influence on the photon--hidden-photon mixing.
This is essentially the fraction in the second line of
Eq.~(\ref{HiddenPhotons:flux:SI})
with $\omega_{\rm p}$ and $\Gamma$ expressed via the more commonly used
complex refractive index $n$ as
$2E_{\gamma^\prime}^2(1-n)=\omega_{\rm p}^2-i E_{\gamma^\prime} \Gamma$
(see, {\it e.g.}, Ref.~\cite{Redondo:2010dp}).
The germanium refractive index data are taken from
Ref.~\cite{Henke1993181}.
Because in our experimental setup the target and the detector are the same, the
efficiency of the system for the expected signal is $\approx$1. 
The X-rays accompanying the photoelectric-like effect will
be thereafter absorbed in the same crystal, so the energy of the particular outgoing 
signal equals the total energy of the incoming hidden photon.
 
   The experimental setup used in this search for solar hidden photons has been described
elsewhere \cite{Horvat:2008ts,Kekez:2008sm,Horvat:2011jh}. Here we only recall that  
the HPGe detector with an active target mass of 1.5~kg
was placed at ground level, inside a low-radioactivity iron box with a wall thickness
ranging from 16 to 23~cm. The box was lined outside with 1~cm thick lead. 
A low threshold on the output provided the online trigger, ensuring that
all the events down to the electronic noise were recorded.
Various calibrated sources have been used to study 
the linearity and energy resolution and, in particular, in the lowest-energy
region mainly a $^{241}$Am source. 
The detector resolution
was about 820~eV for the 13.9~keV gamma-rays and
660~eV for their 3.9~keV escape peak.
Data were accumulated in a 1024-channel analyzer, with an energy dispersion of
63.4~eV/channel and 
with data collection time of $2.38\times 10^7$~s.
In these long-term running conditions, the knowledge of the energy scale is allocated by 
continuously
monitoring the positions and resolution of indium X-ray peaks of 24.14~keV  and 
27.26~keV, which are present in the measured spectra.

   \begin{figure}[ht]
\centerline{\includegraphics[width=98mm,angle=0]{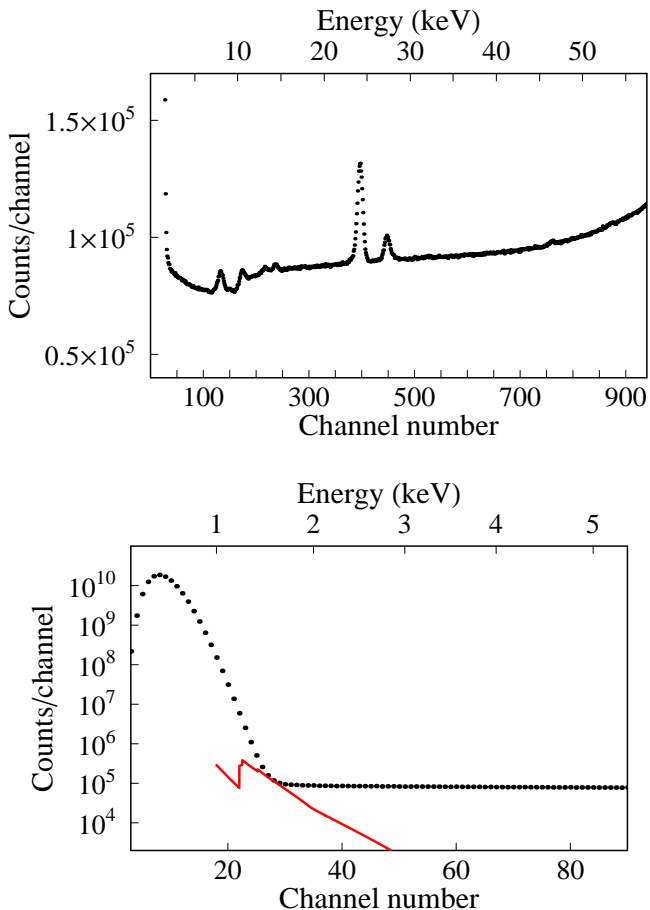}}
               \caption{Top panel: total measured energy spectrum showing also X-ray
peaks from various materials. Bottom panel: low-energy data shown together with
the maximum of expected events due to hidden photon-electron interactions (red line),
corresponding to $E_{\gamma^\prime} \sim 1.6~\mbox{\rm\scriptsize keV}$. }
      \label{2spectra}
   \end{figure}

As can be seen from Fig.~\ref{2spectra}, showing the total energy spectrum, there is
no evidence for any excess of photon-like events due to the hidden photon-electron interactions.
The expected spectrum (red line) has a step at
$E_{\gamma^\prime} \sim 1.3~{\rm keV}$,
corresponding to the threshold in photoelectric cross section on germanium
L electrons. There is no sign of such a step in the measured spectrum
that could positively identify a hidden-photon signal.
Therefore, our evaluation for upper limits on the mixing parameter follows the
most conservative assumption, by requiring the predicted signal
$N_{\gamma^\prime}(k)$ in every energy bin $k$ to
be less than or equal to the recorded counts $N_{\mbox{\rm\scriptsize exp}}(k)$.
Namely, $ N_{\gamma^\prime}(k) \le N_{\mbox{\rm\scriptsize bg}}(k) 
+ N_{\gamma^\prime}(k) = N_{\mbox{\rm\scriptsize exp}}(k)$, where
$N_{\mbox{\rm\scriptsize bg}}(k)$ is the unknown background.
For every fixed $m_{\gamma^\prime}$ we raise the parameter $\chi$ till
the first touch of the predicted spectrum $N_{\gamma^\prime}(k)$
with the measured
spectrum $N_{\mbox{\rm\scriptsize exp}}(k)$ in some channel $k$.
This value of $\chi$ is our upper limit.
Similar approaches have been used elsewhere (see for instance
\cite{Horvat:2011jh,Baudis:1998hi,Morales:2001hj,Angloher:2002in,Angle:2007uj}),
where direct background measurement is not possible
and the signal shape is a broad spectrum on top of
an unknown background spectrum.
Figure~\ref{2spectra} (bottom panel) shows that our experiment is the most sensitive
to the hidden photons of energy around 1.6~keV.
For fixed $E_{\gamma^\prime}(=1.6~{\rm keV})$ and $m_{\gamma^\prime}$, 
the theoretically expected yield of hidden photon-induced events has been
calculated by means of Eq.~(\ref{signal}), where $\chi^4$ is the only free parameter
which is then used to fit the maximal strength of the expected spectrum, marked with red 
line in Fig.~\ref{2spectra} (bottom panel), to the measured one.
For the highest hidden-photon masses under considerations,
$m_{\gamma^\prime}\sim 10~{\rm keV}$,
the energy at which expected spectrum touches the
measured one, shifts from fixed $E_{\gamma^\prime}=1.6~{\rm keV}$
to $E_{\gamma^\prime} >  m_{\gamma^\prime}$.
In order to estimate
a day-night variation of the flux of hidden photons in our experiment (performed in
Zagreb, $\varphi=45^{\circ}45^{'}~{\rm N}$), which is expected due to their travel 
through Earth's mantle\footnote {Earth's mantle is thought to be dominantly oxygen (44.8\%), silicon (21.5\%), and
magnesium (22.8\%) with some (5.8\%) iron and the remainder aluminum, calcium,
sodium, and potassium.} ($2R_{\rm E}\cos\varphi\sim$ 8.9$\times$10$^3$~km in length),
we calculated the absorption under the most conservative assumptions that 
Earth's mantle consists only of iron, and its density is the mean density of the Earth.
It was found that the day-night correction does not affect
our limits on the mixing parameter, for the hidden-photon mass range
displayed in Fig.~\ref{mixing}.
  
The corresponding upper limits on the mixing parameter obtained in this work
are displayed in Fig.~\ref{mixing} together with
the current hidden photon bounds \cite{Redondo:2008aa,Jaeckel:2010ni}. 

In conclusion, we have performed an experiment to obtain the upper limits
on the photon--hidden-photon mixing parameter $\chi$ in the eV to keV
hidden-photon mass range by observing the photoelectric-like process on
germanium atoms inside the HPGe detector impinged by hidden photons coming from the
Sun. 
We have excluded hidden photons with mixing parameters
$\chi > (2.2\times 10^{-13}-3\times 10^{-7})$ in the mass region
$0.2~{\rm eV} \lesssim m_{\gamma^\prime} \lesssim 30~{\rm keV}$.
We then compared our limits on the interaction strength $\chi$ with respect to 
the hidden-photon mass, to that
derived recently \cite{Redondo:2008aa} using helioscope data from the 
CAST experiment \cite{Arik:2011rx},
as well as to those obtained from solar \cite{Redondo:2008aa}
and HB stars \cite{Jaeckel:2010ni} lifetime arguments.
It turns out that our limits in the hidden-photon mass region
from $20~{\rm eV}$ up to $15~{\rm keV}$
are slightly better than those obtained
from CAST laboratory measurement \cite{Redondo:2008aa},
but still somewhat weaker than those obtained from astrophysical considerations
(solar and HB lifetimes). The relevance of our results lies in the fact that
they constitute the best laboratory limits in the said parameter range
obtained to date.

We would like to thank J. Redondo  for useful comments.
The authors acknowledge the support of the Croatian MSES Project No. 098-0982887-2872.

\bibliographystyle{elsarticle-num}

\bibliography{hpM}

\end{document}